\documentclass{elsart}
\usepackage{graphicx}
\usepackage{epsf}
\usepackage{natbib}
\usepackage{amssymb}

\journal{Physica A}

\begin{document}
\begin{frontmatter}

\title{Economic exchanges in a stratified society: End of the middle class?}

\author[ictp]{M. F. Laguna}
\ead{mlaguna@ictp.trieste.it}
\author[if]{S. Risau Gusman}
\ead{srisau@if.ufrgs.br}
\author[if]{J. R. Iglesias *}\corauth[cor]{Corresponding author}
\ead{iglesias@if.ufrgs.br}

\address[ictp]{The Abdus Salam International Center for Theoretical Physics. Strada Costiera 11 (34014) Trieste, Italy.}

\address[if]{Instituto de F\'{\i}sica, UFRGS, Caixa Postal 15051, 91501-970, Porto Alegre, RS, Brazil.}

\date{\today}

\thanks{S. Risau-Gusman acknowledges support from the Centro
Latinoamericano de F\'{i}sica. J.R. Iglesias acknowledges support
from Conselho Nacional de Desenvolvimento Cient\'{\i}fico e
Tecnol\'{o}gico (CNPq, Brazil). The authors acknowledge partial
support from CAPES (Brazil) and SETCYP (Argentina) through the
Argentine-Brazilian Cooperation Agreement BR 18/00.}

\begin{abstract}
We study the effect of the social stratification on the wealth
distribution on a system of interacting economic agents that are
constrained to interact only within their own economic class. The
economical mobility of the agents is related to its success in
exchange transactions. Different wealth distributions are obtained
as a function of the width of the economic class. We find a range
of widths in which the society is divided in two classes separated
by a deep gap that prevents further exchange between poor and rich
agents. As a consequence, the middle wealth class is eliminated.
The high values of the Gini indices obtained in these cases
indicate a highly unequal society. On the other hand, lower and
higher widths induce lower Gini indices and a fairer wealth
distribution.

\end{abstract}

\begin{keyword}
econophysics, wealth distribution, hierarchical systems, Pareto's
law, economic classes \PACS{89.65.Gh \sep 89.75.Fb \sep 05.65.+b
\sep 87.23.Ge}
\end{keyword}

\end{frontmatter}

Several models of capital exchange among economic agents have been
recently proposed\cite{chatter,Matteo,chakra,west1,Bouchaud,solo}
trying to explain the emergence of power law distributions of
wealth obtained by Pareto more than a century ago\cite{Pareto} .
In his original work, Pareto analyzed the distribution of the
income of workers and companies in different countries, and
asserted that in all countries and times the distribution of
income and wealth follows a power law behavior, where the
cumulative probability $P(w)$ of people whose income is at least
$w$ is given by $P(w) \propto w^{-\alpha}$, with $1.2 \leq \alpha
\leq 1.9$ \cite{Pareto}. Recent international empirical data
suggest that Pareto's distribution provides a good fit to the
income distribution of different countries in the range of high
income. Nevertheless, it does not agree with observed data over
low and middle range of income, for which different distributions
were proposed
\cite{Bouchaud,souma,dragu2000,dragu2001a,dragu2001b,clementi}.

Most of the models consider an asymmetric probability that the
poorer agent might be privileged in the exchanges. Increasing the
probability of favoring the poorer agent is a way to simulate the
action of the state or of some type of regulatory policy that
tries to redistribute the resources
\cite{west1,sinha,IGPVA2003,SI2004}. Moreover, almost all these
models consider exchanges of agents picked up either at
random~\cite{IGVA2004}, or following an extremal
dynamics~\cite{IGPVA2003,PIAV2003}. The obtained distribution is a
Gibbs-exponential type in most cases, being the results in good
agreement with real distributions of welfare states \cite{SI2004}.
Other authors have proposed models in which agents have a risk
aversion \cite{chatter,chakra,west1,IGVA2004}. The effect of this
factor on the wealth distribution also gives rise to a
Gibbs-exponential distribution in most cases and shows a power law
behavior in some limits \cite{chakra,IGVA2004}.

All those models have a common point: no correlations between the
wealth of the agents and the probability of interaction between
them are considered. However, the fact that people tends to
strongly interact mainly with others of their own social and
economic class, might be a determinant factor in the wealth
distribution of a population. An example of that is the work of
Inaoka {\it et al.} \cite{inaoka}. They analyzed the exchanges in
Japanese banks and found that big banks have more interactions
between them and with the others than the small ones. A work by
two of the present authors considers this fact by including a
correlation between the success of an agent in their economics
exchanges and his degree of connectivity \cite{seb}.

In this paper we consider a society in which agents are constrained
to interact with others that belong to the same economic class. We
introduce a parameter that establish the maximum difference in
wealth two agents can have in order to interact. This kind of
approach was previously used to study the formation of a public
opinion as the result of social interactions \cite{fab}.

We consider a population of $N$ interacting agents characterized by
a wealth $w_i$ and a risk aversion factor $\beta_i$. We chose as
initial condition for both these parameters a uniform distribution
between $0$ and $1$ \cite{ci}. For each agent $i$, the number
$[1-\beta_{i}]$ measures the percentage of wealth he is willing to
risk. We consider this percentage as an individual fixed parameter
in the whole process. But while the value of $\beta_i$ remains fix,
the value of $w_{i}=w_{i}(t)$ will change as a consequence of the
interactions. At each time step $t$ we first select the two agents
that will exchange resources in the following way: we choose at
random one agent $i$ and, also at random, a second one $j$ that
belongs to the same economic class of agent $i$. It means that agent
$j$ is randomly chosen from the subset of the system for which
$|w_{i}(t)-w_{j}(t)|<u$. The parameter $u$ measures the ``width'' of
the economic class, and determines the number of agents that can
interact with agent $i$ at the time $t$. We also establish that no
agent can earn more than the amount he puts at stake. Then, the
quantity to be exchanged is the minimum value of the available
resources of both agents, i.e., $dw=\min
[(1-\beta_{i})w_{i}(t);(1-\beta_{j})w_{j}(t)]$. Finally, following
previous works we consider a probability $p \geq 0.5$ of favoring
the poorer of the two partners \cite{west1,IGVA2004},

\begin{equation}
\label{eq:sca} p=\frac{1}{2}+f\times\frac{|w_{i}(t)-w_{j}(t)|}{w_{i}(t)+w_{j}(t)},
\end{equation}

where $f$ is a factor going from $0$ (equal probability for both
agents) to $1/2$ (highest probability of favoring the poorer
agent). Thus, in each interaction the poorer agent has probability
$p$ of earn a quantity $dw$, whereas the richer one has
probability $1-p$.

When performing the simulation with these rules, after a transient
the system arrives to a stationary state where the wealth has been
redistributed. We present numerical simulations for a system of
$N=10^4$ - $10^5$ agents, and several values of $f$ and $u$.
Stationary state analysis where made after $t=10^{5}N$
interactions.

We first describe the process of economic exchange between agents.
As we stated before, at $t=0$ each agent receives a wealth in the
interval $[0,1]$. As time evolves, exchanges between agents
generate a redistribution of wealth that, although dependent on
the values of $f$ and $u$, presents some common features. The
first one is that in all the cases the number of people with very
low income increases. As the model is conservative, the resources
of impoverished agents contribute to increase the wealth of other
agents. For low values of $f$ a sharp peak appears for income
equal to zero. This means that a significative fraction of the
population had quickly lost all their resources. The exchange
process is different for high values of $f$ and $u$: a maximum
appears for intermediate wealths and the distribution at long
times seems to be fairer.

\begin{figure}[t]
\centering
\resizebox{\columnwidth}{!}{\includegraphics{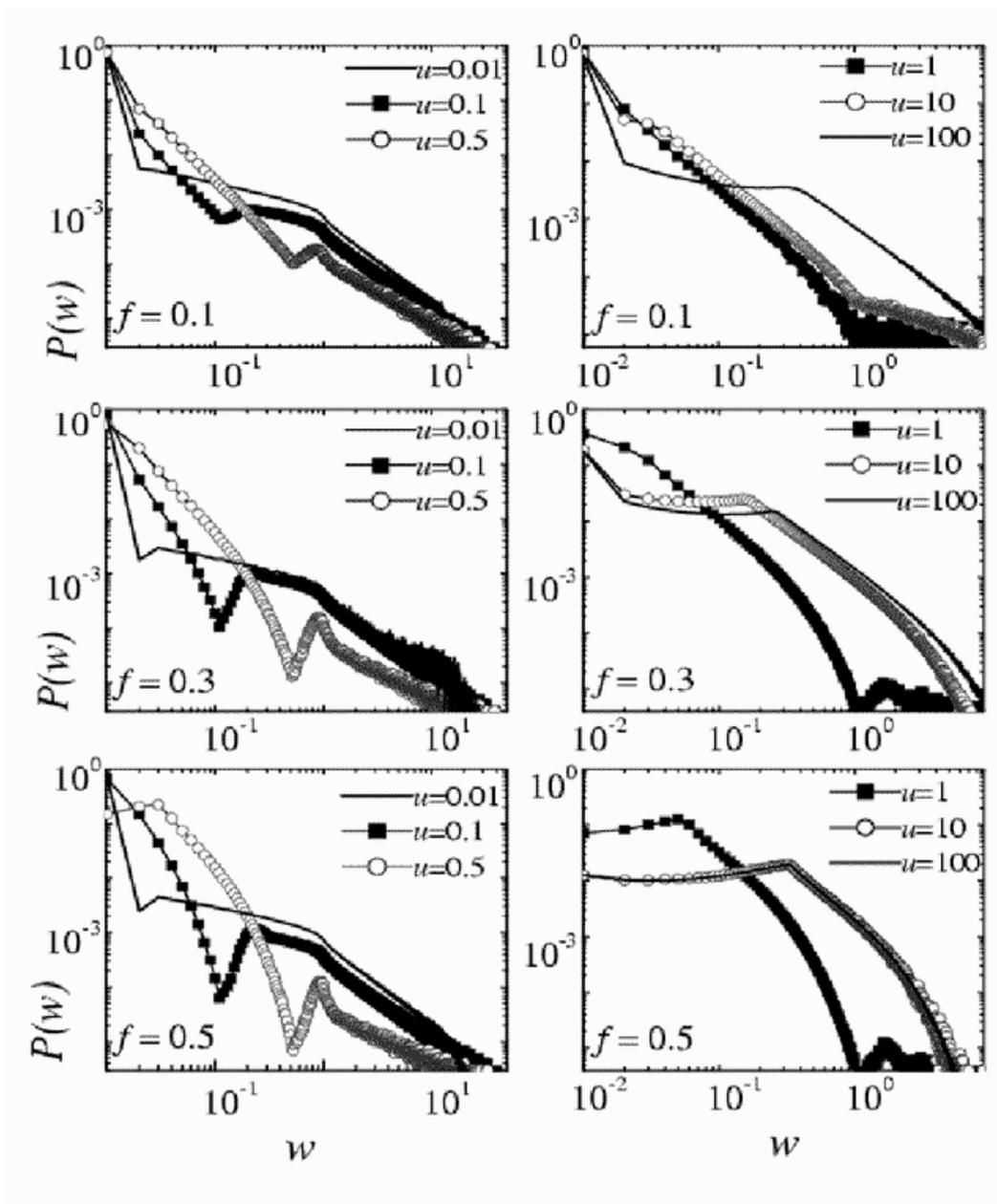}}
\caption{Stationary states for a system of $N=10^{5}$ and several
values of $f$ and $u$. Each curve correspond to an average of $100$
runs. Note that the minimum of the gap corresponds to $w \sim u$. }
\label{fig1}
\end{figure}

A better characterization of this model can be done by analyzing the
stationary states obtained as average of several runs with different
seeds and the same value of the parameters involved. Results are
shown in  Fig. \ref{fig1} for a system of $N=10^5$ agents and $100$
runs for each family of parameters. For very small values of $u$ a
peak in $w=0$ is obtained, as well as a rather flat distribution for
higher values of wealth. This means that very narrow classes prevent
any kind of redistribution. In other words, there is a irreversible
and strong transfer of income from lower to higher classes. The high
peak for income zero is present for low and intermediate values of
$u$ when the probability of favoring the poorer agent is low
($f=0.1$ in the figures). Only for a very high value of $u$ the
distribution is more uniform, with a peak for intermediate values of
$w$. For $f=0.1$ and $0.3$ the peak in $w=0$ disappears for $u
\gtrsim 100$, whereas for $f=0.5$ it happens for $u \sim 1$. But
more interestingly, for all the values of $f$ studied we found an
intermediate range of $u$ in which a formation of two classes is
obtained. One class correspond to agents with a very small wealth,
whereas the other is formed by rich agents. The two classes are
separated by a gap that prohibits further wealth exchanges between
the low and the high classes. This has as a consequence the
elimination of the middle wealth class. The minimum of the gap
happens for $w \sim u$, suggesting that the permitted range of
interaction appears as a scale for the system. This polarization of
the agents in two classes reminds the polarization of opinions that
has been observed, for example, in Ref.\cite{fab} (in spite of the
fact that the rules to change opinions are different that the
present ones to change the wealth). For big enough values of $u$ no
gap is observed, as very few agents can attain a high wealth. The
obtained distribution is similar to the model without restrictions
\cite{IGVA2004}.

\begin{figure}[t]
\centering
\resizebox{\columnwidth}{!}{\includegraphics{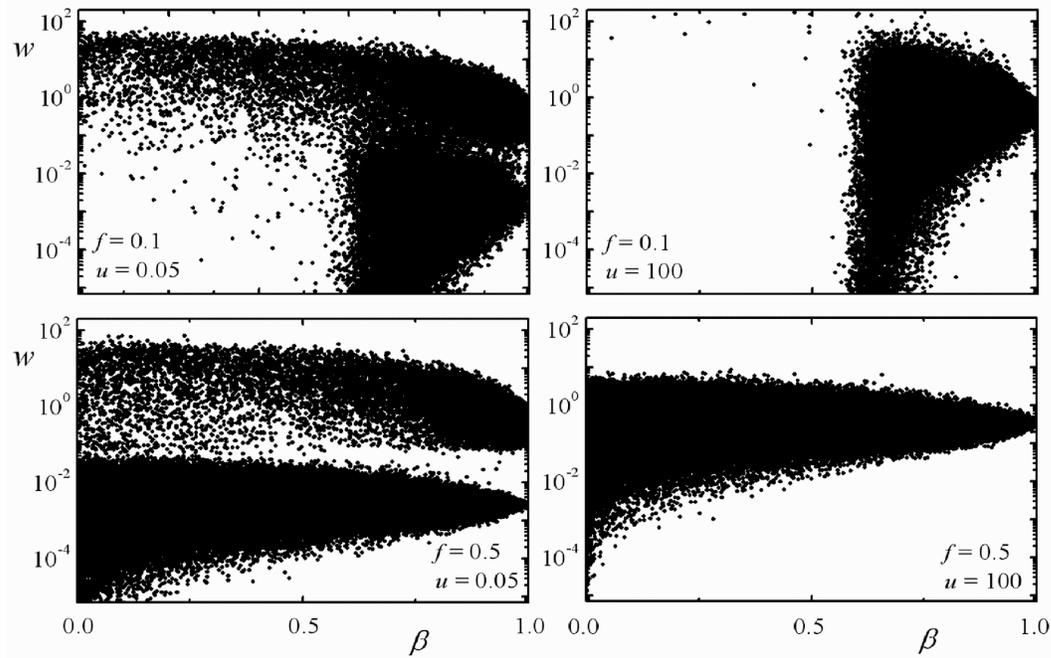}}
\caption{Logarithmic plot of a snapshot of wealth $w$ vs. risk
aversion $\beta$ for a population of $N=10^{5}$ agents. Values of
$u$ and $f$ are indicated in each panel. Each point corresponds to
an agent. For $f=0.1$ agents with very low wealth ($w \sim 0$) are
not visible in the logarithmic plot.}\label{fig2}
\end{figure}

The gap between classes is also observed in the plots of the
correlation between wealth and risk aversion of Fig. \ref{fig2}.
We have represented a particular snapshot of a stationary state
configuration. For low values of $f$ ($f=0.1$ in the plot) the
poorest people have practically zero wealth and, consequently,
they are not seen in the logarithmic plot. For $u=0.05$ a gap
appears in all the range of $\beta$, being much wider in the
region $\beta < 0.5$. This means that agents with a low
risk-aversion can be only very rich or very poor in the stationary
state (the last ones do not appear in the plot). For $u=100$ we
find a distribution in which agents with high values of beta are
in the middle class, as expected for agents that do not risk their
assets, while the richest and poorest agents are those with very
low values of the risk aversion parameter $\beta$. The
intermediate values of $\beta$ assures agents in all the range of
wealth, whereas the poorest agents have the lowest risk-aversion
(again, they do not appear in the plot).

The situation changes if the probability of favoring the poorer
agent is high ($f=0.5$ in the plot). For $u=0.05$ the gap between
the upper and low classes is present in all the range of $\beta$
and can be clearly observed in the plot because the poorest agents
have a low but finite wealth in the stationary state. Agents with
the lowest values of $\beta$ may be in the richest or poorest
stripe of the population. A higher value of the threshold ($u=100$
in the plot) does not present a gap. The richest and poorest
agents are those with very low values of the risk aversion
parameter $\beta$.

Finally, in Table 1 we show the Gini coefficients for the
parameters studied. As can be observed, the highest values
correspond to the intermediate range of $u$, where the gap is
present. Only for high values of $f$ and $u$ we obtain Gini
indexes close to the ones obtained from data of different
countries: while the Gini index that we obtain can go very close
to $1$, the highest values observed in real societies are of the
order of $0.7$.

\begin{table}
\begin{center}
\begin{tabular}{||c||c|c|c|c|c|c|c|c|c||}
\hline
f | u & 0.005 & 0.01 & 0.05 & 0.1 & 0.5 & 1 & 5 & 10 & 100\\
\hline
\hline
0.1 & 0.819 & 0.916 & 0.952 & 0.963 & 0.982 & 0.992 & 0.987 & 0.976 & 0.915 \\
\hline
0.3 & 0.819 & 0.840 & 0.953 & 0.964 & 0.978 & 0.971 & 0.920 & 0.840 & 0.674 \\
\hline
0.5 & 0.818 & 0.915 & 0.950 & 0.974 & 0.962 & 0.932 & 0.690 & 0.488 & 0.472 \\
\hline
\end{tabular}
\end{center}
\caption{Gini coefficients for the three values of $f$ treated in
the article and several values of $u$. The columns in each case
correspond to the different values of $u$ whereas the rows
correspond to the different values of $f$.}
\end{table}

We analyze the behavior of a society in which agents are constrained
to interact with others that belong to the same economic class. We
use a simple model that includes the existence of risk aversion and
an asymmetric probability that the poorer agent might be privileged
in the exchange. Moreover, we introduce a parameter $u$ that defines
the maximum range in wealth for two agents to interact. We studied
the evolution of these systems and found different wealth
distributions depending on the values of the parameters  $u$ and
$f$. For all the values of $f$ studied, we find a range of $u$ in
which the society is divided in two classes separated by a deep gap
that prevents further exchange between them. This gap is related to
the lack of opportunity of the poorer agents to ascend to a class
economically higher. It is important to note that this ``opportunity
gap'' is conceptually different to the poverty gap widely studied,
which has to do with the existence of poor and rich agents. The main
consequence of this wealth redistribution is the disappearance of
the middle class. We remark that we tried to fit the obtained wealth
distributions with Gibbs-like or power law functions. While for low
values of $u$ a power law tail appears over a narrow band of wealth
values, for high values of $u$ an exponential law provides a better
fit. We also calculated the Gini coefficients of the stationary
wealth distributions and find that low and high values of $u$
present a fairer wealth distribution than intermediate values.

\end{document}